# Is There Statistical Evidence that the Oregon Payday-Loan Rate Cap Hurts Consumers?

Assaf P. Oron, Ph.D.
Department of Statistics, University of Washington, August 2009
*assaf "at" uw "dot" edu*

## Executive Summary

1. A recent unpublished manuscript whose conclusions were widely circulated in the electronic media (Zinman, 2009) claimed that Oregon's 2007 payday loan (PL) rate-limiting regulations (hereafter, "Cap") have hurt borrowers.

2. The report's main conclusion, phrased in cause-and-effect language in its abstract - "*…restricting access caused deterioration in the overall financial condition of the Oregon households…*" – relies on a single, small-sample survey funded by the payday-lending industry (PLI). The survey is fraught with methodological flaws.

3. Moreover, survey results do not support the claim that Oregon borrowers fared worse than Washington borrowers, on any variable that can be plausibly attributed to the Cap.

4. In fact, Oregon respondents fared better than Washington respondents on two key variables: on-time bill payment rate and avoiding phone-line disconnects. On all other relevant variables they fared similarly to Washington respondents. **In short, the report's claim is baseless.**

5. There were a few mild observed differences on general economic variables (e.g., respondents' overall future financial outlook). These are far more likely related to the current recession hitting Oregon earlier and harder than Washington, than to the Oregon PL Cap. The report altogether ignores the crisis and its effect on study results, and also repeatedly uses questionable arithmetic to make results appear worse for Oregon.

6. As to certain post-Cap events in Oregon – namely, the PLI's swift and massive exodus from the state and the subsequent reduction in PL activity – the report prefers to repeat the PLI's explanations of these events word-for-word. There is no exploration of alternative causal chains – a half-dozen of which are suggested below. This type of exploration is part and parcel of academic research. In any case, as stated above, the report fails to prove that these events have been detrimental to Oregon borrowers.

## 1. Introduction

Within a decade and a half, payday-loan (PL) stores in most US states have mushroomed from almost nonexistent and downright illegal, to "*more [common]… than McDonald's and Starbucks combined*" (Zinman, 2009).[1] For much of that time, a political battle has been raging between lenders and various consumer and social-justice groups who have opposed this growth. Until a few years ago lenders had held the upper hand, but more recently the tide has been turning, nationwide and state by state. As Zinman (2009) mentions, even President Obama has openly expressed his dim view of the PL phenomenon and its impact upon the poor.[2]

In July 2007, the state of Oregon lowered its PL interest-rate cap from an APR of 528% to 156%, extended the minimum loan duration from 14 to 31 days, and added other consumer-protection measures (hereafter this change, following legislation passed in 2006, will be called "the Cap", using Zinman's terminology). The Cap was followed by an immediate and massive exodus from the state by the payday-lender industry (PLI): over the year 2007 more than two-thirds of Oregon's PL shops had shut down (Graves, 2008).

The report, analyzing a survey of Oregon and Washington PL borrowers immediately before the Cap's enactment (June-July 2007), and a follow-up with some of the same respondents five months later in November-December 2007, claims to have found a causal connection between the Cap and Oregonians' financial well-being. This connection is summarized succinctly in the abstract, as quoted above in my Executive Summary ("…*restricting access caused deterioration in the overall financial condition of the Oregon households*…"), and is spelled out in more detail later on:

> *"The five-month results above suggest that the Oregon Cap reduced the supply of credit for payday borrowers, and that the financial condition of borrowers (as measured by employment status and subjective assessments) suffered as a result."* (Zinman, 2009, p. 13)

The suggested causal chain of events is schematically drawn in Fig. 1. **Declaring a causal connection based on statistical analysis is a bold step** – especially a multi-stage connection like that depicted in Fig. 1. The proper method to find such a connection is a randomized controlled experiment – and a multi-stage connection probably requires several experiments. By contrast, the results of field surveys and observational studies performed on society are usually framed in terms of associations and not in cause-and-effect language. This is not a mere formality: in a well-run controlled experiment we can eliminate or neutralize the impact of all possible causes except for the examined one. In other study types – including surveys and observational studies – such isolation of the effect of interest is typically not feasible. We cannot eliminate the possibility that other factors have caused the observed signals; this is known as confounding (Friedman et al., 2007, Ch. 1-2).

---

[1] My first critique of this study referred to the original October 2008 version. Since then, two updated versions were posted by Dr. Zinman. The first update (12/08) was a very minor change. The more recent version (3/09) had more substantial changes to the style and organization, but the general analysis, discussion and message have remained almost identical. Some important interim data were dropped from the March 2009 version, and I will refer to these data later on. However, in general the critique addresses the latest version, as is customary.

[2] I will refrain from explaining here what PLs are. The nature of these loans and their benefit or damage to society are highly interesting to discuss at this unique point in American history, but the discussion is mostly tangential to the article's main focus and is therefore omitted here (there will be a brief discussion PL profitability in Section 3).

Under certain conditions it may be acceptable to reach a level of confidence that allows one to declare causality without a controlled experiment. However, typically the evidence needs to be overwhelmingly strong and broad, coming from many different independent studies. The Surgeon General committee examined 7,000 scientific articles before concluding that smoking causes lung cancer without the "smoking gun" of a randomized experiment (Surgeon General, 1964).

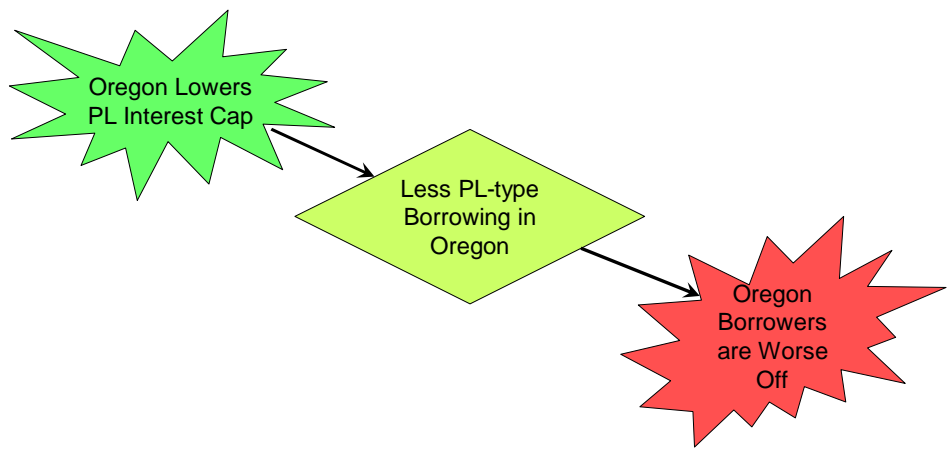

**Figure 1:** The main causal chain of events that, according to Zinman (2009), connects the Cap to Oregon borrowers being worse off. More detail will be shown in Fig. 4 (Section 3).

The study of the impact of a policy change, or of any other sudden, drastic event affecting society, is known as "quasi-experimental analysis" (Campbell and Ross, 1968). This type of study examines "before vs. after" differences, and whether they should be attributed to the policy change. One way to partially disentangle the change's effect from background historical trends, is to look at similar neighboring regions (localities, states, nations) in which the change did not take place. In the survey reported by Zinman (2009), the state of Washington, which is generally considered to most resemble Oregon in its geography, society and economy, was chosen. The study can thus be roughly described as a "case-control quasi-experiment", with Oregon being the "quasi-case", Washington the "quasi-control", and the Cap being the "quasi-treatment".

This could be a plausible analysis frame. Nevertheless, Washington is *not* Oregon, and the survey in question is *not* a controlled experiment. It is impossible to remove all confounding factors that make Washington's society and economy different from Oregon. The burden of examining whether – for a particular study – the two are similar enough for one of them to act as a reasonable "quasi-control" to the other, lies upon the researchers and not the audience. In any case, the language used to describe any results from such a study should be much more careful and nuanced than the statements in Zinman (2009).

Besides using careful language, a responsible analysis of uncontrolled surveys and observational studies must include some meaningful discussion of other potential causes for any observed effects (again I refer the reader to Ch. 1, 2 and 19 of the Friedman et al., 2007 basic-statistics textbook). This discussion is largely absent from the report, which frames the survey as if it were a randomized experiment.

With this key point made, it is still worthwhile to inspect whether the report's main claim still stands, albeit on a weaker footing. The first stage of the causal chain in Fig. 1 – i.e., the association between the Cap and decreased PL use in Oregon - does not seem to be in much doubt, and in fact many of the Cap's proponents may see it as a favorable outcome.[3] The more critical question, therefore, is whether the Cap, directly or via decreased PL use, is indeed clearly associated with deterioration in the financial condition of Oregon's pre-Cap PL borrower population. This will be dealt with immediately below in Section 2, which is the longest section of this critique. Section 3 examines the public events surrounding the PLI's exit, and their interpretation in Zinman (2009). The critique ends in Section 4, with conclusions and some recommendations.

---

[3] That being said, whether the causal connection for the post-Cap reduction in PL activity is as simple as the author would have us believe is an open question, which will be dealt with later in Section 3.

# 2. Are Oregonians Worse Off?

## 2.1 The Sample

The 2008 election season has once again demonstrated the potential of sample surveys. When survey questions are simple and clear-cut, and when the sample is representative, a survey of several hundred citizens can provide a general indication of where the entire public stands. However, if the sample is *not* representative, all bets are off and the survey's results may be completely meaningless or even misleading (Friedman et al., 2007, Ch. 19). Therefore, a major question is whether the discussed survey's sample is representative of the borrower population.[4]

According to Zinman (2009), a sample of 17940 persons was randomly drawn from all Oregonians and Washingtonians who took a PL from one of four large companies. Out of this sample, the survey firm obtained complete pre-Cap interviews from only 1040 borrowers - 7% of the Oregon sample and 5% of the Washington sample. This is a rather small proportion. The author does not clarify whether this second selection was random too. Since the 1040 responders were split into precisely 520 from each state, it is quite likely that surveyors stopped upon reaching a specific quota – and so, respondents may come from a sub-population which was easier to contact. There are additional reasons why the social and financial traits of responders and non-responders may be quite different. This is known in general as **non-response bias**; it is a problem that plagues survey research to varying degrees. The non-response bias of the pre-survey sample is not discussed in the report.

To see how a bias in the sample can distort inference, let us observe the results of a single survey question: whether the borrower was late in paying any bill during the past three months (information taken from Zinman's Table 1). Out of 520 pre-Cap Oregon respondents, 430 (83%) responded in the affirmative. If the 520 respondents are a representative sample, then with reasonable confidence we can assume that 83% is fairly close to the true proportion of late-bill payers among the entire Oregon borrower population of the four participating PL companies. However, if the sample is biased, and if we cannot find a reliable way to estimate the bias, then all we can do is to calculate the range within which the true proportion lies. Assuming that the original Oregon sample had about 7500 borrowers[5] s, the true proportion of late-bill payers in the initial sample could be anywhere from 6% (430 out of 7500) to 99% (7410 out of 7500) – rendering the reported proportion of 83% almost meaningless.

---

[4] Another major factor distorting survey results is the wording, framing and ordering of questions. We cannot examine this issue because the report does not provide the complete survey questionnaire texts.
[5] This is based on extrapolation from the 7% response rate reported, since the exact information was not given.

To make matters worse, 1040 borrowers were not the ultimate sample from which Zinman (2009) draws his conclusions. Only 873 of the 1040 (441 in Oregon and 432 in Washington) agreed to participate in the follow-up survey. This is yet another source for bias, since agreement to follow-up is clearly not a random selection. From among them, the survey firm only interviewed 200 Oregonians and 200 Washingtonians – again, hinting at a rigid quota. Zinman does not specify whether this last selection of a 400-person sample - only 2.2% of the initial random sample - was also random. Moreover, some respondents declined to answer certain questions. For example, in answer to the post-survey's late-bill question, 146 out of 196 Oregonians (74%) reported having paid bills late at least once over the past three months. But the true proportion in the only sample which can be plausibly seen as random – i.e., the original sample - could be anywhere from 1.9% (146 of 7500) to 99.3% (7450 of 7500).

At least one aspect of the final sample selection process was definitely *not* random: anyone who had their phones disconnected during five months between pre- and post-surveys –73 of the 441 Oregonians (16.6%) and 103 of the 432 Washingtonians (23.8%) – was not contacted (this information was taken from Zinman, 2008; the 2009 version omitted it). Since it is plausible that these borrowers were on average in a worse situation than those still using a phone line (e.g., they were *far* more likely to have paid a bill late; perhaps even their phone bill), and since their proportion in Washington is significantly larger, their elimination from follow-up – though perhaps inevitable – should cause an overall bias. This bias would likely make Oregon respondents seem worse off in the post-survey (relative to Washington) than they really are.[6]

Table 1 below summarizes the sampling stages described here, and my concerns about them. Since Zinman (2009) provides rather incomplete information about the sampling process, it is possible that he received the sample data from the PLI after the fact, without having been involved in survey design and oversight. Indeed, the author's description of the data source, and of his relationship with CCRF (the PLI body that funded the survey), does not rule out this possibility. If so, the entire study is automatically placed under serious ethical and practical doubts (this point will be discussed at greater length in Section 4).

Reserving our doubts for the moment, we will (for the remainder of Section 2) still assume that the sample is fairly representative of the two states' borrower populations, and that data integrity had been maintained. Even then, the sample size of 400 borrowers, split into two groups, interviewed both before and after the Cap (with some questions answered by as few as 330 respondents) is rather on the small side. Needless to say, almost any result from such a sample needs to be described using very careful and qualified language.

---

[6] Zinman (2009) does attempt to address the selection bias in going from the 1040 pre-sample to the 400-person post-sample via several weighting scheme. However, his methods seem rather questionable (see Appendix B).

| Stage | Resulting Sample Size | Type | Concerns and Open Questions |
|---|---|---|---|
| "All" borrowers from 4 PL companies | ? | Proxy for population | Which companies? What is the population size? Were duplicate entries managed? |
| Initial sample | 17940 | Random | Why the large initial sample, if so few were ultimately used? |
| Pre-survey sample (June-July 2007) | 1040 (520 each state) | ? | **How were the 1040 chosen? In any case, there is a non-response bias** |
| Agreed to participate in post-survey | 873 (441 OR, 432 WA) | Volunteers | Non-response bias |
| Had same phone line working during post-survey | 697 (368 OR, 329 WA) | Socio-economic Attrition | **Selection bias (WA "surviving" borrowers are a relatively stronger subset)** |
| Post-survey sample (November-December 2007) | 400 (200 each state) | ? | **Why not interview all 697? How were the final 400 chosen?** |

**Table 1:** Summary of the sampling process for the survey as reported by Zinman (2008, 2009).

## 2.2 Using Washington as a Control, and the Historical Context

As stated above, choosing Washington for a surrogate quasi-control is generally plausible. In normal times, comparing trends in these two states over a five-month period might be a fairly safe affair (within the limitations of an observational study, of course). However, as far as the economy is concerned, we are now in anything but "normal times". Here is what the report has to say about the economic background:

> "*Oregon and Washington are neighboring states that were on similar economic trajectories at the time of the surveys: both states had experienced 4 consecutive years of employment growth, and both states forecasted a flattening of employment rates for the latter half of 2007 (Oregon Office of Economic Analysis 2007; Washington Economic and Revenue Forecast Council 2007).*" [Zinman, 2009, pp. 8-9]

This description depicts a background of economic tranquility and (somewhat dampened) prosperity, which is perhaps how the cited state planning bureaus had indeed seen things in early 2007. But since survey respondents based their answers on *real-time experience* and not on prior economic forecasts by their state governments, let us refresh our knowledge from a mid-2009 perspective, and place the study period - the second half of 2007 - in its correct economic and historical context.

These months were in fact an early stage of what is widely seen as the nation's worst economic crisis since the Great Depression. In mid-2007 the housing bubble was already bursting across most of the country; financial markets were extremely jittery from July 2007 onwards; and the impact upon "Main Street" was already tangible enough that by 2007's end there was a rare agreement between the Bush White House and the Democratic Congress leadership, on the need for an economic stimulus package. In fact, it has been determined that by late 2007 the nation was already officially in recession. Moreover, there is also a general agreement that the crisis was first felt among economically vulnerable populations – i.e., the constituency containing most PL borrowers. Therefore, one can safely assume that the economic situation of quite a few of Zinman (2009)'s survey respondents was affected by the crisis during the study period.

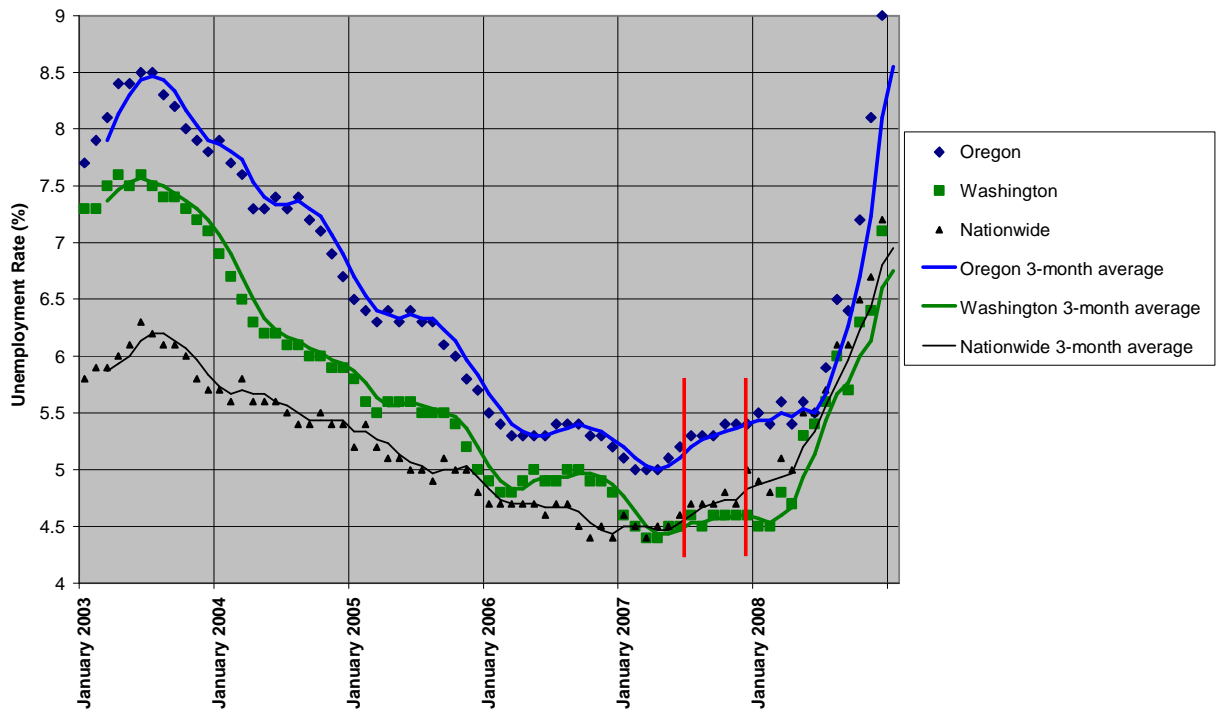

**Figure 2:** Unemployment-rate trends from 2003 through 2008, for Oregon (blue), Washington (green) and the entire United States (black). Symbols indicate monthly values, while solid lines are based on moving averages of the 3 most recent months. The two red vertical lines indicate the approximate time of Zinman (2009)'s pre- and post-surveys. Most data were downloaded from the Federal Bureau of Labor Statistics website in November 2008, with updates made until January 2009.

Let us examine more closely how Oregon and Washington were related to the national context, and to each other, in late 2007. Fig. 2 presents the same key measure mentioned in the quoted passage above: state jobless rates from 2003 through 2008. The pre- and post-survey dates are indicated with two vertical red lines. As can be seen, both Oregon (blue) and Washington (green) suffered badly from the dot.com-9/11 economic crisis: unemployment in both was higher than the national average (black). Subsequently both states were late in riding the housing bubble, and late in its bursting. Eventually, they have caught up with the nation in falling into the present recession.

However, there the similarities end. Oregon's post-9/11 recession, in terms of jobless rate, was a full percentage point worse than Washington's – even though the two were nearly even with ~5% unemployment during the year 2000. More directly relevant to our study, Oregon's housing-bubble-related job growth was far less enduring than Washington's. From spring 2007 onward, Oregon's unemployment rate has been getting progressively worse in tandem with the national trend. In contrast, throughout 2007 Washington was defying the trend and enjoying an economic "Indian Summer" of sorts.

Why the Oregon and Washington trajectories differed during the study period is an open question. Perhaps the impact of Oregon's most populous neighbor – California, where the housing bubble and its burst were quite dominant – was felt upon Oregon suppliers and contractors. Or perhaps it is some inherent weakness and vulnerability in Oregon's economy, evident from its unemployment trend throughout the present decade (as of June 2009, Oregon unemployment was 12.1% - among the highest in the nation and nearly 3% worse than Washington's).

The precise answer is immaterial to our discussion. All we need to do here is examine whether Washington can be used as a reasonable quasi-control for general economic measures over the study period; **the answer to this question is no**. In other words, given recent economic history and the Oregon and Washington trends, it is preposterous to blame any relative weaknesses in the Oregon sample's *general economic variables* upon the Cap. Yet this is precisely what Zinman (2009) presents as the main piece of evidence for Oregon borrowers being "worse off".

## 2.3 Analyzing Survey Results

### a. Results

The discussion above is summarized in Fig. 3. At this point, it already seems that the report's original causal conclusion rests upon very weak foundations. Yet, the question remains whether the 200 Oregonians in the sample *were indeed worse off* than the 200 Washingtonians, as Zinman (2009) claims. Data with potential answers to this question are presented in Zinman (2008)'s Tables 2 and 3 (the 2009 report had some items dropped; see below). The line items in these tables can be roughly divided into three types of personal economic activity: 1. short-term loan usage patterns, 2. general economic condition (including employment status), and 3. ability to meet ongoing financial obligations. Short-term loan usage patterns, while interesting, are not in themselves a direct indicator of being better or worse off – which is Zinman (2009)'s main claimed result. In fact, as argued in the introduction, decreasing borrowers' reliance upon high-cost loans has been seen as a *desirable* result by Oregon's policymakers.

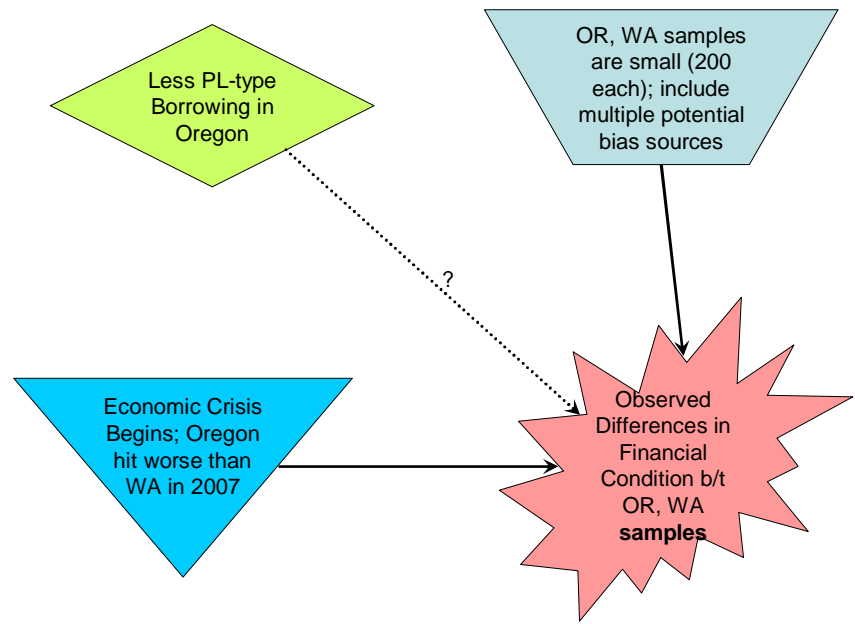

**Figure 3:** Potential causes to the financial condition of borrowers in the Oregon sample (relative to Washington), in light of my analysis in Section 2.1.

As to overall economic conditions, given the national and regional background just discussed, such measures - on which the Oregon sample fared mildly worse in general - mean little to our bottom line. They are related to the recession, first and foremost.

The most relevant group of questions, therefore, is the third and last one, namely: how well did borrowers in the sample handle **specific ongoing financial obligations** in late 2007, compared with mid-2007. I counted five questions in the third group, and my re-analysis for them using Chi-Square tests based on relative-risk analysis is presented in Table 2.[7] A relative-risk greater than 1 between Oregon and Washington indicates that the Oregon sample fared worse. The corrected p-value conveys the relative strength of the evidence.

---

[7] This measure differs somewhat from the "difference in differences" measure used by Zinman (2009); explanations why this approach is preferable are found in Appendix A. Regarding the choice of 5 relevant questions out of the 18 presented by Zinman (2008): in order to enable readers to evaluate my judgment on that matter, all remaining questions in the survey are listed on Table C1 in Appendix C.

In general, results are similar to Zinman's raw analysis (Zinman, 2008, Column 5 of Table 2). On most variables there is no significant difference between the two samples. The only significant signal (question C4, in bold) **is in Oregonians' favor:** the number of Oregonians paying bills late at least once over the 3 months prior to survey dropped significantly (from 168 to 146, out of 196), while the corresponding Washington number remained nearly flat (from 149 to 150, out of 197; exact counts were reconstructed from Zinman's table). This translates to a 14% lower adjusted relative risk for the Oregon sample, with a p-value of 0.04, corrected for multiple comparisons (see further below). **Interestingly, between Zinman (2008) and Zinman (2009) seven questions were dropped from his Table 2, including this one.**

| Variable | Relative Risk (OR vs. WA) | Chi-Square Statistic | Raw P-value | Corrected P-value |
|---|---|---|---|---|
| C1: Not having an Account Protection* | 0.96 | 0.11 | 0.74 | 0.93 |
| C2: Overdraft/Bounce at least Once | 1.03 | 0.09 | 0.76 | 0.76 |
| C3: Overdraft/Bounce Twice or More | 1.13 | 0.60 | 0.44 | 0.82 |
| **C4: Late Bill at least Once** | **0.86** | **6.91** | **0.009** | **0.04** |
| C5: Late Bill "Frequently" | 0.86 | 0.61 | 0.44 | 0.90 |

**Table 2:** The five line items from Zinman (2008)'s Table 2 which are directly related to borrowers' financial health, re-analyzed using the Chi-Square test. The first item was changed from "having an account protection" to "not having an account protection", in order to align its direction with the others (i.e., more means Oregon faring worse). The p-value correction (rightmost column) accounts for looking at five questions, none of which was pre-determined to be of more interest than any other.

An additional key measure related to handling financial obligations was discussed in Zinman (2008, p. 13-14), and mentioned above in Section 2.1: the rate of phone disconnects. Of the 873 respondents agreeing to participate in the follow-up, 16.6% of Oregonians and 23.8% of Washingtonians had their line disconnected during the study period's five months. This difference is clearly significant in Oregon's favor (Chi-Square statistic: 6.73, p-value: 0.009). Moreover, since this information is not subject to the last sample selection stage and is based on a sample of 873 borrowers rather than only 400, it is arguably more representative of the general borrower population than the data in Table 2. The entire passage containing this information was edited out of Zinman (2009).

**b. Methodological Concerns**

The author fails to guard against the fallacy of **multiple comparisons:** suppose for the moment that we made the survey on two groups of robots programmed to give completely random answers. From basic probability theory we know that even with such essentially meaningless noise, the chance of yielding a faux "significant difference" between the groups on at least one question – *any* question - would increase as the number of questions increases. Conversely, increasing the number of questions, coupled with a willingness to accept any signal that comes our way without taking care of multiple comparisons, will turn any study into little more than a "fishing expedition". Standard approaches to correct for multiple comparisons are roughly equivalent to multiplying the p-value. Thus, with 20 questions, each single p-value would be multiplied by a factor of up to 20 (see e.g. Miller, 1981).

Not only does the report ignore multiple comparisons; the author also combines several original survey questions to create new comparisons, thus engaging even more aggressively in a "fishing expedition". Consider for example the measure presented as key evidence that Oregonians have fared worse (see, e.g., the passage quoted in my p. 2). This measure (last line item of Zinman (2009)'s Table 3 and my Table C1, Appendix C) examines whether respondents were unemployed, *or* partially employed, *or* saw their financial situation deteriorate recently, *or* expect it to deteriorate in the future. This item exhibits three fatal flaws: it is strongly recession-related, it represents little that makes sense or that can be plausibly attributed to the Oregon Cap – and last but not least, its significance is inflated by not correcting for multiple comparisons.

Finally, the author attempts various ways of re-weighting the sample via propensity scores – and then treats these processed versions more seriously than the raw-data version. Absent solid information to the contrary, the most reliable snapshot is provided by raw sample responses. In our case, reliability is severely limited given the small sample and the questionable sampling process, as discussed above in Section 2.1. Propensity scores are a speculative way to fix these problems. The scores themselves are estimates, and therefore subject to uncertainty; an uncertainty neglected by in the report.

More fundamentally, key theoretical assumptions for using propensity scores are grossly violated in this dataset, thus adding more doubts and biases rather than resolve existing ones. In particular, the assumption that sample attrition is unrelated to the effects of interest is clearly violated – since attrition by losing one's phone service is probably related to financial condition (Zinman, 2009, footnote 25, p. 9).[8]

---

[8] A more detailed discussion of Zinman (2009)'s propensity-score approach is found in Appendix B.

## 3. What "Restricted" the "Access"?

Even though by now the report's main claim is thoroughly debunked, for reasons that will become apparent soon I would like to revisit the first leg of the claimed causal connection– namely, Oregon's post-Cap drop in PL borrowing (recall Fig. 1). Let us assume, as before, that the sample is sufficiently representative; then the survey does indeed indicate that there was such a reduction. Only about half of the Oregon pre-sample reported recent PL borrowing in the post-sample, compared with 79% of the Washington sample; this is a strongly significant signal by any measure.

According to Zinman (2009), the direct cause for the contraction of PL business in Oregon is not the Cap itself, but rather the *sudden supply destruction* caused by payday-lender exodus – first and foremost, the exit of large national corporations with dozens of statewide branches each:

> *Ex-ante, these* [i.e., the Cap – AO] *were plausibly binding restrictions on payday lenders that typically charged at least $15 per $100 for two-week loans pre-Cap, since there does not seem to be any compelling evidence that payday lenders make excess profits (Flannery and Samolyk 2005; Skiba and Tobacman 2007). Fixed costs, loan losses, and related risks can account for market rates of 390% APR. Ex-post, the binding nature of the Cap is evidenced by payday lenders exiting Oregon."* [Zinman, 2009, p. 6]

Zinman (2009)'s complete causal explanation why Oregon borrowers are "doing worse", based on this and previously-quoted passages, is illustrated in Fig. 4. It focuses on the supply destruction as the sole cause for reduction in PL borrowing. However, during the three months before the post-survey, there were well over 100 PL shops in Oregon: substantially less than before, but still quite available. In metropolitan Portland's poor neighborhoods, for example, the decrease in PL shops might boil down to extending the trip to the nearest shop from half a mile to a mile (roughly speaking, a reduction by a factor of 4 increases distances between neighboring shops by a factor of 2). In view of this – given that PL price was dramatically lowered by the Cap – why wouldn't borrowers be just as willing to make that trip? If the product was so good for them, why didn't they crowd the remaining stores now that it is even cheaper?[9] A complete answer to this question requires knowledge I do not possess.[10] My educated guess, in view of the short five-month time frame, is that the drop in PL borrowing reported in the survey may simply reflect initial confusion from the sudden disappearance of most PL outlets.

---

[9] The author does mention this paradox (Zinman, 2009, p.14), but does not follow through on it to revisit his own assumptions.

[10] Some of this information could have been easily available, had the PLI survey presented by Zinman (2009) cared to ask questions such as "*why* didn't you take a PL recently?"

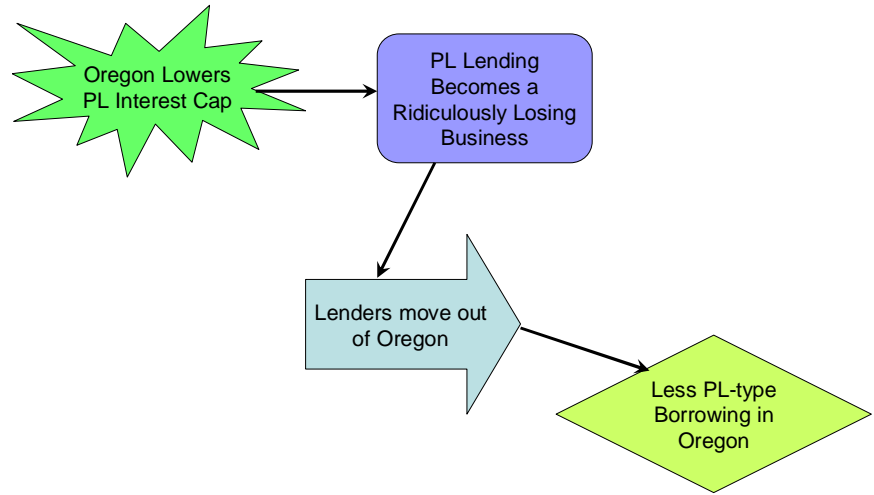

**Figure 4:** The causal link between the Cap and decrease in PL-type borrowing in Oregon, according to Zinman (2009), in greater detail than shown in Fig. 1.

Here are a few more alternative explanations. First, it is not outlandish to assume, in view of the increased local media attention to the issue, that public perceptions of PLs had become more negative in Oregon during 2006-2007. Therefore, many Oregonians have perhaps become more reluctant to stigmatize themselves as PL borrowers, or unwilling to risk falling into what the majority of Oregon's policymakers saw as a debt trap. Another possible cause for decreased borrowing is that social organizations – the same ones who lobbied to pass the Cap law – engaged in more extensive outreach efforts to would-be PL borrowers, helping them find lower-cost substitutes. Again, these are informed speculations rather than definitive answers. But it is Zinman's report – supposedly an academic study - that should have included a wealth of such speculations, and a proper survey questionnaire should have included questions examining them.

Moreover, Zinman (2009) completely ignores payday lenders' free choice. According to him, lenders are either automatons or natural phenomena: once the Cap was enacted, it was inevitable that they leave the state because business became impossible. What Zinman's argument really boils down to, is taking the PLI at its word that it is not as profitable as one might think.

However, the author himself admits in his introduction that the PL profitability debate is far from settled. My humble, non-economist view is that the proof is in the pudding: an industry does not become a boom market boasting 20% annual growth for an entire decade – unless it is very profitable indeed. Furthermore, across the nation, even after this phenomenal growth and overcrowding of PL shops, most PLs are still made at the maximum allowed APRs. These two pieces of evidence indicate that the PLI operates quite far above its lower profitability bound.

Furthermore, they suggest that the PLI's primary negotiation over rates is not with its customer base but with politicians and regulatory agencies.[11]

Continuing this line of thought, it is rather plausible that most PL companies – especially the deep-pocketed national corporations that were actually the first to fold – *could have* stayed, tried to adapt their Oregon business and continue making money from it, rather than shut it down so quickly. But why should a national corporation stay in hostile, Capped Oregon, when next door in California or Washington, and elsewhere, it can still take advantage of higher rates and a more favorable political climate? In short, a far more likely explanation for the PLI's exit than the one given in the report, is that Oregon is simply paying the price of being first. Taking this explanation a bit farther, some of these companies may have made the calculation that making the Oregon exit as sudden, painful and widely publicized as possible may pay off in the longer term, as a deterrent to other states from following Oregon's example.

In summary: the report's author repeats the PLI's version of events almost word for word, and fails to explore alternative explanations. It is worth noting that the PLI survey apparently did not bother to ask borrowers who stopped taking PLs, why they did so. In view of the complex web of possible causes illustrated in Fig. 5, such information would have been quite valuable.

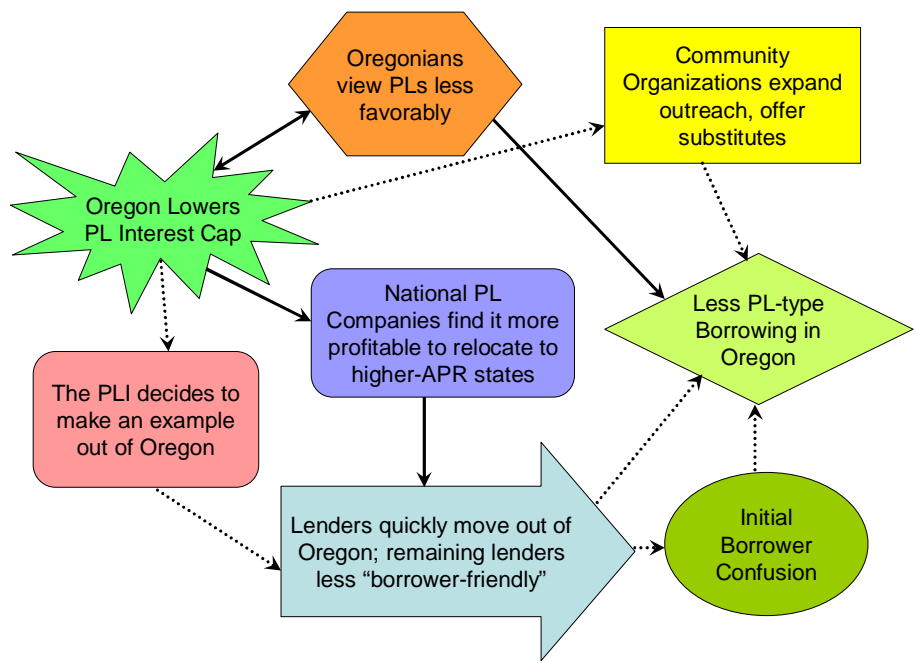

**Figure 5:** Similar to Fig. 4, only here the causal chains are the ones suggested by my analysis and by an online media search. The more plausible links per my judgment are marked by solid lines.

---

[11] The PLI is certainly not at fault for the huge growth of this captive credit-constrained population – but just like the rest of the greater PL debate, this question is tangential to the report discussed here.

# 4. Conclusions and Recommendations

I reiterate that the conclusions put forth by Zinman (2009) are utterly baseless, for the following major reasons:

1. The article draws its information from a single survey, which is not nearly enough material to proclaim any causal connection, surely not in the strong language used by the author. Furthermore, this survey was not a controlled experiment.

2. The survey's sampling process was tortuous, and several critical sampling stages in this process are not properly explained. Those stages which are explained give rise to multiple sources of bias, making the sample probably not representative of borrower population. Moreover, the final sample – 200 Oregonians and 200 Washingtonians – is quite small.

3. The study period – five months from mid-2007 to late 2007 – occurred during the slide into the current recession, which is a strong source of confounding signals for any economic study. The author ignores the economic crisis altogether.

4. Oregon's economy has suffered from the economic crisis earlier and more severely than Washington – and this difference was already apparent during the study period. Thus, "pre vs. post" comparisons on general economic measures of the Oregon sample should generally look worse than similar comparisons on Washington sample – without any need to invoke other causes such as the Oregon Cap. The only "significant" difference pointing against the Oregon sample was found on a combination of general economic measures.

5. The only two relevant, specific and significant signals regarding personal financial condition coming from the survey – on-time bill payment and the ability to retain phone service – actually point in favor of the Oregon sample.

6. The methods used for calculating differences and their significance are severely flawed; for example, the effect of multiple hypothesis testing has been completely neglected.

7. The causal chain argued in the report – the Cap leading to lender exit, leading to less PLs taken, leading to economic effects – is far from being the only plausible explanation for the chain of events. In fact, it is not very plausible, and more likely explanations exist. The author fails to explore them.

8. There is a suspicious similarity between the terminology, arguments and warnings of PLI executives following the Cap's enactment, and the terminology, arguments and conclusions of the report.

Even if one accepts the report's propensity-score modified versions of the data rather than the raw results, and even if one is willing to recognize the "unemployed *or* partially employed *or* had a rough half-year *or* fearful of the future" combination as a Cap-relevant indicator (and again, there is no reason to accept either) – it still means that out of about a dozen indicators of financial condition, only one or two point against the Oregon sample and the rest are indeterminate. Yet this incredibly weak signal, produced thanks to immense efforts of selective logic and creative arithmetic, is rhetorically used by the author as a causal proof beyond doubt that the Cap made Oregon borrowers worse off. This is very poor scholarship.

Given the report's funding source, one issue which should inevitably be examined is the role of **industry-sponsored research.** Industry-sponsored research has already tainted much of the medical sciences, to such a degree that the editors of the Journal of the American Medical Association (JAMA) wrote an editorial in Spring 2008 with the strongly worded title "*Impugning the Integrity of Medical Science: The Adverse Effects of Industry Influence*" (DeAngelis and Fontanarosa, 2008). This editorial ended with a list of 11 new requirements for medical articles published in JAMA, on top of previous restrictions – all designed to neutralize industry influence upon the way studies are analyzed and reported. Or course, medical research is far more tightly monitored than economics, in terms of analysis methods and standards; but even there, once industry money and influence are allowed to dominate independent research, the truth is harmed and the public interest ill-served.

The JAMA editorial paid special attention to the phenomenon of "*ghost writers*": industry employees or contractors who actually draft the initial article, but are not credited as authors of the final manuscript, and sometimes not even mentioned in the acknowledgments. I trust Dr. Zinman that he is indeed the text's sole author, as he declares in footnote 6. However, it is not completely clear *who designed the survey*. We know that the survey was paid for by the PLI; were there also "*ghost surveyors*" from the industry who actually designed the survey and decided what would be asked and how, what questions to leave unasked, and (possibly) what data to censor out after the fact? The lack of complete documentation and explanation of all the survey's sample-reduction steps points towards this explanation. If so, then the "ghost surveyors" should be included as co-authors of the report, since a survey's design and questionnaire play a critical role in any conclusion drawn from it.

I also object to the use of certain partisan terms in a way that lends them undue academic credibility. The report substitutes the term *"market rates"* for maximum regulatory rates (see Section 3 above); this is factually wrong. The author also describes the Cap as "*access restriction*" – a term appearing with minor variations in the article's title, four times in its abstract and at least ten times in the text. Of course, the Cap, whatever its effect, was <u>not</u> an access restriction. Oregon consumers are still free to go to any PL store offering these loans; the restrictions were on providers' fees and not on anybody's access. Arguably, the Cap is an access *expansion* for consumers, since reducing the maximum rate makes PLs more affordable. The term "access restriction" attaches a negative, totalitarian connotation to the Cap. It is no surprise that PLI executives would use it; academic researchers should use more factual terms.

In terms of research, what the PL field needs are more studies by truly independent academics, using independent and impartial (or public-interest) funding sources. It is lamentable that so few such researchers have found the issue interesting enough to pursue. There are many deeper questions about PL economics – for example, what market model would adequately describe their dynamics; or what are the underlying processes contributing to their booming success, and whether these processes point to social and economic problems that need to be addressed using other means. These questions remain largely unexplored save for partisan studies by industry-related or consumer-advocacy groups, or studies using woefully inadequate economic models that have been thoroughly debunked by the current crisis. As we have so recently learned, leaving "the hidden economy of the poor" to fester in the dark - or to be prodded only by those profiting heavily from it and by their allies - is a losing strategy.

## Conflict of Interest Statement

The author of this critique has performed this consultation independently and *pro bono*, at the request of Columbia Legal Services (CLS), a non-profit organization offering legal assistance to vulnerable Washington State populations. There was no compensation to the author, whether direct or indirect. The author had studied a PL-related question in the past, on behalf of the University of Washington Statistical Consulting service at the request of the Seattle Post-Intelligencer (Oron, 2006). The author has no direct financial stake in PL businesses or in commercial interests related to this market, and has received no editorial input or censorship from CLS.

## Acknowledgments

I am grateful to Mr. Bruce Neas of the Columbia Legal Services, for asking me to examine Dr. Zinman's article, and thus to become involved once again in this important social and economic problem. Thanks to Professor June Morita of the statistics department, University of Washington, for her helpful comments on the sampling method; and to Prof. Thomas Richardson of the same department, for immensely useful discussion and information.

# References

Agresti, A., 2002. *Categorical Data analysis, 2nd Edition.* Wiley.

Campbell, D.T. and Ross, H.L., 1968. The Connecticut Crackdown on Speeding: Time-Series Data in Quasi-Experimental Analysis. *Law & Society Review*, Vol. 3, No. 1, pp. 33-54.

DeAngelis, C.D. and Fontanarosa, P.B., 2008. Impugning the Integrity of Medical Science: The Adverse Effects of Industry Influence. *Journal of the American Medical Association*, 299, 1833-1835 (editorial article).

Friedman D., Pisani R. and Purves R., 2007. *Statistics, 4th Edition.* W. W. Norton (note: one can also use earlier editions for the material discussed here).

Graves, B., 2008. Oregon's payday lenders all but gone. *Oregonian,* July 6, 2008. Retrieved November 2008 from the Oregonian website (www.oregonlive.com).

Miller, R.G., 1981. *Simultaneous Statistical Inference, 2nd Edition.* Springer-Verlag, New York.

Oron, A.P., 2006. Easy Prey: evidence for race and military related targeting in the distribution of Pay-Day loan branches in Washington State (revised version, March 2006). Consulting report, Statistical Consulting Service, University of Washington.

Richardson, T.R., 2008. Estimation of the relative risk and risk difference. Department seminar, Department of Statistics, University of Washington, 2008.

Surgeon General's Advisory Committee on Smoking and Health, 1964. *Smoking and Health.* United States Public Health Service.

Zinman, J., 2008. Restricting consumer credit access: household survey evidence on effects around the Oregon rate Cap. Working paper; revised version, December 2008. Retrieved January 23, 2009 from Zinman's Dartmouth webpage (http://www.dartmouth.edu/~jzinman/)

Zinman, J., 2009. Restricting consumer credit access: household survey evidence on effects around the Oregon rate Cap. Working paper; revised version, March 2009. Retrieved July 20, 2009 from Zinman's Dartmouth webpage.

# Appendices

## A. "Difference in Differences" vs. Relative Risk

Zinman's approach to analyze survey responses was what he calls "difference-in-differences" (DD), which is simply

$$DD = \Delta p_{OR} - \Delta p_{WA} = p_{OR,after} - p_{OR,before} - (p_{WA,after} - p_{WA,before}),$$

with the *p*'s indicating observed proportions of "yes" responses to a given survey question. This method is most suitable for variables that are not restricted, i.e. they can take any value. However, proportions are restricted between 0 and 1 (or, equivalently, between 0% and 100%), creating potential distortions. For example, quite often we would not view a change from 1% to 6% the same way as a change from 45% to 50%; the former is arguably more drastic. However, the DD approach treats them as equal. Sometimes this is appropriate; sometimes not, depending upon the research context. Moreover, the normal approximation justifying the p-values reported for DD calculations begins to break down when the number of observed "yes" or "no" responses is only in the single digits. This does occur in several questions, most prominently in the question whether responders find it hard to get short-term loans (last line-item in Zinman's Table 2; item A7 in my Table C1, Appendix C). In Washington's pre-survey, only 8 respondents answered "yes" - rendering the DD significance calculations overly optimistic.

An arguably safer standard approach to "yes-no" type survey data is **a Chi-Square test of homogeneous association** on the three-way table generated by the division of responses into OR vs. WA and the two sets of "yes" vs. "no" answers to any question (one set before, one set after; see e.g. Agresti, 2002, Section 6.3). Zinman does not provide complete data to construct the data's three-way tables, but the details in his Table 2 are sufficient to reconstruct the data margins and perform a somewhat simplified version of the Chi-Square test. In order to calculate the null distribution for this test, I used the measure of **relative risk** (Richardson, 2008)**.** Risk here is a synonym for probability, so the relative risk (RR) is

$$RR = \frac{RR_{OR}}{RR_{WA}} = \frac{p_{OR,after}/p_{OR,before}}{p_{WA,after}/p_{WA,before}} = \frac{p_{OR,after} \, p_{WA,before}}{p_{WA,after} \, p_{OR,before}}.$$

With RR, a doubling in "yes" response rate from 5% to 10% will be equivalent to a doubling from 10% to 20% - and much stronger than a 5% increase from 45% to 50%. Both the RR-based Chi-Square test and the difference-in-proportion (Zinman's "DD") test are acceptable measures under certain limitations. With large enough samples, close enough to a 50%-50% split between "yes" and "no", and when Oregon's and Washington's initial rates are similar, the two should yield similar results. Otherwise, deciding which test is preferable depends upon the research context, and in case of doubt it is a good idea to examine both.[12] I found the RR approach to be more appropriate, since when the two state samples had different baseline levels it makes more sense to compare each state's proportional change, rather than the absolute change.

## B. Correcting for Sample "Attrition"

Zinman (2008,2009) does not stop his analysis with simple calculations based on the raw data from 400 respondents. Rather, he re-analyzes the data three times – in all cases, producing conclusions somewhat more favorable to the PLI's cause (i.e., showing Oregonians as worse off in the post-survey). For example, the late-bill rate signal, significant in Oregon's favor in the raw sample, becomes insignificant after the author assigns each respondents a weight designed to make the sample of 200 in each state more "representative" of the original sample, or to make the Washington sample demographically more "similar" to the Oregon sample, using propensity-score weighting (columns 6 and 7, respectively, of the report's Tables 2-4). A key assumption in the weighting schemes is that the difference between the 200 respondents and the remaining 320 in each state sample, is not related to financial outcome (Zinman 2009 footnote 25, p. 9) – an assumption clearly violated here (see above, Section 2.1).

There are additional reasons not to prefer a re-weighted analysis over the raw one. First, the sample is too small to support such manipulations with high confidence. Second, there are no off-the-shelf standard methods yet for such re-weighting, and therefore the door is open to keep tweaking weights until the desired signal is obtained. Finally, given the sample selection process and the nature of the questions, there is no safe and reliable way to "salvage" a larger and more representative picture using the smaller sample's data. Specifically, the group of 640 borrowers (320 in each state) excluded from the post-sample is in fact a collection of three groups with different characteristics:

1. Those who refused to participate in the post-survey (79 in Oregon, 88 in Washington) – a source of non-response bias;

2. Those (from among the remaining group) who had lost their phone line during the study period (73 in Oregon, 103 in Washington) – a source of economically-related bias;

---

[12] With both methods, the test needs to be modified in case all responses are "yes" or "no".

3. From the remaining group, a quota excluded to yield exactly 200 from each state (168 in Oregon, 129 in Washington) – no bias if this last exclusion was random (if it was *not* random, then the entire survey and its analysis are worthless).

In light of this, it would have been far better to try and interview all 697 respondents who agreed to a post-survey and still had a working phone line, and thus eliminate the last of the three groups. It would have also been worthwhile to try and locate the ones who have lost (or changed) their phone service – and thus minimize the size of the second group, which is the most liable to cause outcome-related bias. In any case, assuming the final 200-person quota selection stage was random, the third (and largest) group should be equivalent to the post-sample. But Zinman (2009)'s propensity-score weighting does not consider this three-group structure; rather, it lumps all three together – while any conscientious attempt to use propensity scores should have modeled each group separately.

## C. Table of Other Survey Questions

| Variable | Zinman's Analysis | Comments |
|---|---|---|
| A1: Took a Payday loan | Less in Oregon | Agreed; but this is not a relevant indication of financial health |
| A2: Took an Auto-Title loan | No Signal | Not a relevant indicator |
| A3: Took a Credit-Card Cash Advance | No Signal | Not a relevant indicator |
| A4-6: Various combinations of A1-3 and C2-5 | Less in Oregon | Combinations that mean little; signal is dominated by PL-taking gap |
| A7: Harder to get Short-Term Loans? | More in Oregon | Signal much weaker when using standard Chi-Square approach (p=0.10 before correction), and when using multiple-testing correction. Also quite possibly affected by economic crisis. |
| B1: Unemployed | No significant signal | Hardly related to the question of interest (effects of the Cap); recession-affected. |
| B2: Unemployed/Partially-employed | No significant signal | Hardly related to the question of interest (effects of the Cap); recession-affected. |
| B3: Financial situation worse? | No significant signal | Subjective and vague; recession-affected. |
| B4: Future financial outlook worse? | Oregon weakly more (significant at 10% level, uncorrected) | Subjective and vague; no multiple-comparison correction; recession-affected. |
| B5: Yes to any of B1+B4 | More in Oregon | Combination meaningless and not called for; responses strongly recession-affected; corrected p-value using relative-risk is 0.12. |

**Table C1:** All remaining questions from Zinman (2009)'s Tables 2 and 3, which were not included in my Table 2. The prefix A indicates questions related to borrowing patterns; B to general economic situation.